\documentclass[twocolumn,showpacs,aps,prl]{revtex4}

\usepackage{graphicx}
\usepackage{dcolumn}
\usepackage{amsmath}
\usepackage{epsfig}
\RequirePackage{xspace}
\begin{document}

\begin{flushright}
{AMES-HET 02-05}\\
{BNL-HET-02/12}\\
\end{flushright}

\def\op{{\cal O}}
\def\lsim{\mathrel{\lower4pt\hbox{$\sim$}}\hskip-12pt\raise1.6pt\hbox{$<$}\;
}
\def\Dd{\psi}
\def\pp{\lambda}
\def\ket{\rangle}
\def\BAR{\bar}
\def\xba{\bar}
\def\fa{{\cal A}}
\def\fm{{\cal M}}
\def\fl{{\cal L}}
\def\ufs{\Upsilon(5S)}
\def\gsim{\mathrel{\lower4pt\hbox{$\sim$}}
\hskip-10pt\raise1.6pt\hbox{$>$}\;}
\def\ufour{\Upsilon(4S)}
\def\xcp{X_{CP}}
\def\ynotcp{Y}
\vspace*{-.5in}
\def\etap{\eta^\prime}
\def\bfb{{\bf B}}
\def\fd{r_D}
\def\fb{r_B}
\def\ed{\eta_D}
\def\eb{\eta_B}
\def\hatA{\hat A}
\def\hatfd{{\hat r}_D}
\def\hated{{\hat\eta}_D}

\def\uglu{\hskip 0pt plus 1fil
minus 1fil} \def\uglux{\hskip 0pt plus .75fil minus .75fil}

\def\slashed#1{\setbox200=\hbox{$ #1 $}
    \hbox{\box200 \hskip -\wd200 \hbox to \wd200 {\uglu $/$ \uglux}}}
\def\slpar{\slashed\partial}
\def\sla{\slashed a}
\def\slb{\slashed b}
\def\slc{\slashed c}
\def\sld{\slashed d}
\def\sle{\slashed e}
\def\slf{\slashed f}
\def\slg{\slashed g}
\def\slh{\slashed h}
\def\sli{\slashed i}
\def\slj{\slashed j}
\def\slk{\slashed k}
\def\sll{\slashed l}
\def\slm{\slashed m}
\def\sln{\slashed n}
\def\slo{\slashed o}
\def\slp{\slashed p}
\def\slq{\slashed q}
\def\slr{\slashed r}
\def\sls{\slashed s}
\def\slt{\slashed t}
\def\slu{\slashed u}
\def\slv{\slashed v}
\def\slw{\slashed w}
\def\slx{\slashed x}
\def\sly{\slashed y}
\def\slz{\slashed z}

\title{
\vskip 10mm
\large\bf
Getting $\beta - \alpha$  without penguins
}

\author{David Atwood}
\affiliation{Dept. of Physics and Astronomy, Iowa State University, Ames,
IA 50011}
\author{Amarjit Soni}
\affiliation{ Theory Group, Brookhaven National Laboratory, Upton, NY
11973}

\date{\today}

\begin{abstract}
Oscillation effects in $B^0\to K_S D^0$ and related processes are
considered to determine $\delta\equiv \beta-\alpha + \pi = 2 \beta 
+ \gamma$.  We suggest
that $D^0$ decays to CP eigenstates used in concert with inclusive $D^0$
decays provide a powerful method for determining $\delta$ cleanly {\it
i.e.} without any complication from penguin processes. The CP asymmetry is
expected to be $\lsim$40\% for $D^0$ decays to non-CP eigenstates and
$\lsim$80\% for decays to CP eigenstates. This method can lead to a fairly
accurate determination of $\delta$ with $O(10^8-10^9)$ $B$-mesons.
\end{abstract}


\maketitle

The two asymmetric B-factories have made remarkable progress in
determining one of the angles ($\beta$) of the unitarity triangle; the
world average now stands at $sin 2\beta^{WA} = 0.78 \pm 0.08$
\cite{higuchi,raven}. This is in very good agreement with the expectations
from the Standard Model (SM), $sin 2\beta^{SM} = 0.70 \pm 0.10$
\cite{ciuch}. However, considerable amount of theoretical input has to be
used to deduce $sin 2\beta^{SM}$ and progress in reducing the theory error
is likely to be rather slow.  Thus, methods that determine the angles
without the uncertainties of hadronic matrix elements are crucial in
testing the CKM paradigm\cite{cabib} to an increasing degree of accuracy
in an effort to search for CP-odd phase(s) due to physics beyond the SM.

In the SM, CP violation is controlled by only one CP-odd phase. 
Therefore, different decays which measure the same angle of the
unitarity triangle (UT) may give inconsistent results if physics
beyond the SM is present. Likewise
other 
apparent failure of unitarity of
the CKM matrix,
such as the failure of the UT to close, 
would also indicate new physics. Beyond the phase $\beta$,
the determinations of $\alpha$ and $\gamma$, therefore,
provide key SM
tests.

Two extensively studied methods for determining $\alpha$ already exist,
via $B \rightarrow \pi \pi$ \cite{gron} and $B \rightarrow \rho \pi$
\cite{snyder}. In these approaches, in addition to some experimental
difficulties, 
considerable theoretical input is essential as theses modes receive
large QCD-penguin as well as some electro-weak penguin (EWP) 
contributons. While efforts at these methods should certainly
continue, it is also very important that, in our drive towards
precision, we develop methods
that require no theoretcal assumption and therefore have
zero theory error. The key point is that effect
of beyond the SM CP-odd phase(s) on B-physics may be quite small
so any residual theory error on the determined unitarity
angles may mask the effect of new physics and thwart
experimental searches.

In this work we wish to report on our study of a method to extract
$\delta \equiv \beta-\alpha+\pi = 2 \beta + \gamma$ 
that uses interference between $b \rightarrow u$ and $b
\rightarrow c$ {\it tree graph exchanges only}; no penguin contribution,
strong or EW, or any theoretical assumption is involved. 
Given that $\beta$ is already well measured, this method
is very
effective in determining $\alpha$ ``cleanly'', i.e. without QCD
complications. 
In addition, this method
can also be used to simultaneously extract
$\beta$, allowing a crucial check against the value of $\beta$ determined
with the 
$B \rightarrow J/\psi K_s^0$ approach \cite{higuchi,raven,bigi_sanda}.
As mentioned before, a difference in the two determinations of $\beta$
may then be an indication of new physics.  The basic idea behind the
method has already received 
some attention \cite{grontwo,branco,kayser,sanda,dunietz}.  
We extend and complement these earlier studies in several ways so that it
becomes now a powerful approach to determine $\alpha$, and possibly
$\beta$, without any complication from penguins.

In principle, a comparison of time dependent CP asymmetry measurement in
$B^0 (\bar B^0) \rightarrow K_S D^0$ with that in $K_S \bar D^0$ suffices
to give $\delta$ \cite{branco,sanda}. In practice, though, as has already been
noted previously, flavor tagging of $D^0 (\bar D^0)$ appears extremely
difficult \cite{kayser,atwood}. Semi-leptonic tags suffer from very
serious background from prompt B-decays, $B \rightarrow l \nu X _C$;
therefore, here we will not consider the possibility of semi-leptonic tags
further. Hadronic tags of $D^0$ (say via $D^0 \rightarrow K^- \pi^+$)
receive appreciable corrections from doubly-Cabibbo-suppressed decays of
$\bar D^0$. As in the case of $\gamma$ extraction with $B^{\pm}$
\cite{atwood}, this interference can be used to our advantage in
determining $\delta$ as Kayser and London (KL) have discussed
\cite{kayser}.

In this letter, we would like to highlight at least two additional methods
which will be shown to have great practical importance in extracting
$\delta$.  First of all, $\delta$ may be determined if $D^0$
decays to CP eigenstates (CPES) are observed, provided both CP=$+1$ and
CP=$-1$ states are used. Although we find that while neither this CPES
method, nor the CP non-eigenstate (CPNES) method of KL\cite{kayser}, can
separately provide an especially sharp determination of $\delta$, a great
improvement is achieved if both approaches are used together because both
data sets depend on a common set of parameters in the amplitude.  
Secondly, we will generalize these methods from single final states to
inclusive sets of final states. In this way we can use the entire
observable hadronic branching ratio of the $D^0$ greatly enhancing the
statistical power.  Finally we will briefly discuss methods whereby
ancillary information constraining $\delta$ may also be obtained. The
methods which we describe share with KL\cite{kayser} the feature that the
amplitude parameters are overdetermined and therefore a value of $\beta$,
in addition to $\delta$, may also be extracted from the same data
providing a valuable comparison to $\beta$ obtained from $B\to J/\psi
K_S$.

%To establish the formalism we 
Consider now  
the case where $B^0(t)/\xba B^0(t)\to K_S D^0$, $K_S \xba D^0$
followed by the decay
$D^0/\xba D^0\to F$; 
%In the above we assume that both $D^0$ and $\xba D^0$ channels
%contribute and 
$F$ denotes an inclusive set of states $F=\{f_i\}$ and in
general $F\neq\xba F$.  For example, the set $\{f_i\}$ may range over
states of different particle content (e.g. $K^-+n\pi$) or different points
in phase space~\cite{index_note} (e.g. each $f_i$ is a point on the
$K^-\pi^+\pi^0$ Dalitz plot)  or a combination of both.
For each $f_i$ the four relevant amplitudes are:

\begin{eqnarray}
\fa_1(f_i)&\equiv& \fa(\xba B^0\to K_S [D^0\to f_i])=A
\nonumber\\
\fa_2(f_i)&\equiv& \fa(B^0\to K_S [\xba D^0\to f_i])=A\fd e^{+i\ed}
\nonumber\\
\fa_3(f_i)&\equiv& \fa(\xba B^0\to K_S [\xba D^0\to f_i])
=A\fd\fb e^{+i(\ed+\eb-\gamma)}
\nonumber\\
\fa_4(f_i)&\equiv& \fa(B^0\to K_S [D^0\to f_i])
=A\fb e^{+i(\eb+\gamma)}
\label{fouramps_def}
\end{eqnarray}

\noindent where, 
we have adopted the Wolfenstein\cite{wolf} representation
of the CKM matrix, and
without loss of generality, we can choose the strong
phase convention so that $\fa_1=A$ is real. The quantity $\fd$ is the
ratio $\left | \fa(\xba D^0\to f_i)/\fa(D^0\to f_i)\right |$ which we will
assume is known from the study of $D^0$ decay. The strong phase
$\ed(f_i)=arg\left ( \fa(\xba D^0\to f_i)/\fa(D^0\to f_i)\right )$ we will
assume to be not known apriori.  Likewise the parameter $\fb$ and the
strong phase $\eb$ given by $\fb e^{i\eb}=e^{-i\gamma}\fa(B^0\to K_S
D^0)/\fa(\xba B^0\to K_S D^0)$ are also assumed to be not known apriori. Note
that $\{\fd$, $\ed$, $A\}$ depend on the state $f_i$ while $\{\fb$,
$\eb\}$ are independent.

The  time dependent decay rates for this decay is:

\begin{eqnarray}
\begin{array}{l}
%2{d\over d\tau}
2 \Gamma(B^0/\xba B^0(t)\to K_S F)
\\
=e^{-|\tau|}(X(F)+bY(F)\cos(x_B\tau)-bZ(F)\sin(x_B\tau))
\end{array}
\label{time_dependence}
\end{eqnarray}

\noindent where $\tau=\Gamma_B t$ and $x_B=\Delta m_B/\Gamma_B$
while
$b=+1$ for $B(t)$ and $b=-1$ for $\xba B(t)$.  
Defining $\fa(f_i)=\fa_2(f_i)+\fa_4(f_i)$ and
$\xba\fa(f_i)=\fa_1(f_i)+\fa_3(f_i)$, the coefficients $X$, $Y$ and $Z$
in Eqn.~(\ref{time_dependence}) are given by
$2X(F)$
=
$
\sum_i ( \left|\fa(f_i)\right|^2+\left|\xba\fa(f_i)\right|^2  )
$;
$2Y(F)$
=
$
\sum_i ( \left|\fa(f_i)\right|^2-\left|\xba\fa(f_i)\right|^2 )
$
and
$Z(F)$ 
=
$
\sum_i Im  ( e^{-2i\beta} \fa(f_i)^*\xba \fa(f_i) )
$.
We can expand these quantities in terms of eqn.~(\ref{fouramps_def})
and obtain

\begin{eqnarray}
X(F)  
&=&
\big ( (1+\hatfd^2)(1+\fb^2)/2
\nonumber\\
&+&2R_F\fb\hatfd\cos(\hated-\gamma)\cos\eb \big )
\hatA^2
\nonumber\\
Y(F)  &=&
-
\big ( (1-\hatfd^2)(1-\fb^2)/2
\nonumber\\
&-&2R_F\fb\hatfd\sin(\hated-\gamma)\sin\eb\big )
\hatA^2 
\nonumber\\
Z(F) &=&
\big (
R_F\fb^2\hatfd\sin(2\alpha+\hated)
-R_F \hatfd\sin(2\beta+\hated)
\nonumber\\
&+&\hatfd^2\fb\sin(\eb-\delta)
-\fb\sin(\eb+\delta)
\big)
\hatA^2
\label{XYZ_incl}
\end{eqnarray}

\noindent
where $\hatA^2=\sum_i
A^2(f_i)$, $\hatfd^2 = (\sum_i A^2(f_i)\fd^2(f_i))/\hat A^2$ and $R_F
e^{i\hated} = (\sum_i A(f_i)\fd(f_i)e^{i\ed(f_i)})/(\hat A\hatfd)$.

The corresponding quantities for $\xba F$ are given by 
$X(\xba F)(\eb$, $\ed$, $\gamma) = X(F)(-\eb$, $-\ed$, $\gamma)$;  
$Y(\xba F)(\eb$, $\ed$,$\gamma) = -Y(F)(-\eb$, $-\ed$, $\gamma)$ 
and 
$Z(\xba F)(\eb$, $\ed$,$\gamma) =Z(F)(-\eb$, $-\ed$, $\gamma)$ 
assuming that there is no additional CP violation in $D^0$ 
decay~\cite{CPK_note}.

Initially we will assume that $\beta$ is well determined.  Let us now
consider the special case where $F$ consists of CPES with eigenvalue
$\sigma=\pm 1$.  In this case, the modes add coherently and so $R_F=1$,
$\hatfd=1$ and $\hated=0$ or $\pi$ for $\sigma=+1$, $-1$ respectively.  
The three observables $X(F)$, $Y(F)$ and $Z(F)$ thus depend on the four
parameters $\{\hatA$, $\fb$, $\eb$, $\delta\}$.  If we have the two data
sets, for $\sigma=+1$ and for $\sigma=-1$, then there are five independent
observables (note that $Y(\sigma=+1)+Y(\sigma=-1)=0$) determining the same
four parameters and so the system is overdetermined and one may solve for
$\delta$.

Some examples of CP=$-1$ final states \cite{pdb} include $K_S\pi^0$
(BR=1\%), $K_S\eta$ (0.35\%), $K_S\rho^0$ (0.6\%), $K_S\omega$ (1.1\%),
$K_S\eta^\prime$ (0.9\%) and $K_S\phi$ (0.4\%) giving a total of about
4.4\%. CP=+1 final states include $K_Sf_0$ (0.3\%), $\pi^+\pi^-$ (.07\%)
and $K^+K^-$ (.21\%) for a total of 0.6\%. For each of the modes with a
$K_S$ one can construct a mode of the opposite CP by changing the $K_S$ to
a $K_L$. 
%\cite{KL_can_do}. 
We can also change the $K_S$ which arises from the $B^0$ decay
i.e. in $B^0\to K_S D^0$ (which we refer to as the fast kaon) to a $K_L$.  
Switching the fast kaon to $K_L$ changes $\eb\to\eb+\pi$ and thus gives
the same information as switching the slow kaon (i.e. the kaon arising
from $D^0$ decay). It should be emphasized that, in this
instance, including the final states both with $K_S$ and with $K_L$,
increases the number of observables and, as mentioned above,
enables the system of equations to become soluble. 
This is in contrast, for example, with the case of $B^0\to
J/\psi K_S$ versus $B^0\to J/\psi K_L$ where switching the kaon merely
improves statistics but does not provide additional independent
observables.

We can extend this CPES method to consider inclusive final states.  If $F$
is defined in a CP invariant manner (eg. $F=K_S+X$, BR=21\%) the resultant
observables will be similar to the pure eigenstate case. Here again
$\hatfd=1$ and $\hated=0$ or $\pi$ but $R_F$, which measures the purity of
$F$, will not be $1$. The 3 observables are thus dependent on 5 
parameters $\{\hatA$, $\fb$, $\eb$, $\delta$, $R_F\}$.  As before, we can
obtain a solution by changing the fast $K^0$ to a $K_L$ and/or changing the
slow $K^0$, in the case where $K_S\in F$.  Again,  these $K^0$ changes
will lead us to 5 observables.

In ~\cite{kayser} KL studied the special case where $F$ consists of a
single quantum state which is a CPNES (e.g. $f=K^-\pi^+$). Then 
$R_F=1$ but $\{\hatA$, $\fb$, $\eb$, $\delta$, $\ed\}$ are not known.  If
we take the point of view that $\beta$ and all the relevant $D^0$
branching ratios are well determined then as discussed in~\cite{kayser}
there are 6 observables $\{X(f)$, $Y(f)$, $Z(f)$, $X(\xba f)$, $Y(\xba
f)$, $Z(\xba f)\}$ determining these 5 parameters and so the system
is overdetermined; therefore, one can extract $\delta$. Furthermore, as
in~\cite{kayser} one can also take the point of view that $\beta$ is a
free parameter and solve for both $\beta$ and $\delta$ from the same six
observables. In this context, 
%we want to add  
%that, 
as in the CPES case,
taking the fast kaon to be $K_L$ (rather than $K_S$) provides 
6 more independent observables dependent on the same parameters
rendering the system even more overdetermined.

There is a great advantage to combining the CPES and CPNES methods
above since the parameters involved in the CPES case are a subset of those
for a CPNES. Thus, combining information from CPES and CPNES methods can
increase the number of observables to nine or eleven depending on whether
one or both CP eigenvalues are included, respectively. 
Indeed, if also the fast $K_L$ is taken with the CPNES 
then the number of observables increases to seventeen.
The number of
parameters, of course, stays the same, i.e. five (or six if we also
include $\beta$ as an unknown). Thus, not only there is enough information
but in fact there is considerable degree of redundancy to 
%allow for a
%meaningful solution of 
solve for the unknown parameters.

Likewise considering several CPNES can enhance the degree of over
determination. For each CPNES we add, we have six new observables but
introduce only one new parameter ($\ed(F)$) giving a net gain of 5.  
Indeed there are several candidate modes: $K^-\pi^+$ (branching ratio
3.8\%), $K^-\rho^+$ (10.8\%), $K^{*-}\pi^+$ (5.0\%), $K^{*0}\pi^0$
(3.1\%), $K^{*-}\rho^+$ (6.1\%) and $K^{*-}a_1^+$ (7.3\%)  giving a total
36\%.  In this method one would have to separate the quasi two body modes
from the broad resonances (eg $K^-\rho^+$) making it somewhat difficult.

Generalizing the CPNES case to inclusive states should provide the
most statistically powerful data to determine $\delta$. For instance, the
inclusive $D^0\to K^-+X$ has a branching ratio of 53\%~\cite{pdb}. In this
case, we have the general case of eqn.~(\ref{XYZ_incl}) and so six
observables $\{X(F)$, $Y(F)$, $Z(F)$, $X(\xba F)$, $Y(\xba F)$,
$Z(\xba F)\}$ are determined by the six parameters $\{\hatA$, $\fb$,
$\eb$, $\fd$, $\ed$, $\delta\}$ and so the system can be solved with some
discrete (8-fold) ambiguities.

This may be improved in two ways.  Firstly, one can segregate the set $F$
into several subsets. Thus each additional set $F$ provides six new
observables but introduces only two new parameters ($\hat \ed$ and $R_F$)
giving a net gain of 4.  For instance, a substantial fraction of $K^-+X$
is made up of the exclusive state $K^-\pi^+$ (4\%) together with the
inclusive (in the sense that these modes depend on phase space variables)
states $K^-\pi^+\pi^0$ (13.9\%), $K^-\pi^+\pi^+\pi^-$ (7.5\%),
$K^-\pi^+\pi^-\pi^0$ (10.0\%) and $K^-\pi^+\pi^+\pi^-\pi^0$ (4.0\%)  
giving a total branching fraction of about 40\%. Another approach would be
to divide the $K^-+X$ into separate bins according to the energy of the
$K^-$ in the $D^0$ frame. This would approximate the above since a higher
energy $K^-$ would tend to be associated with fewer pions; then 
one would not have to identify the content of the $X$ state. Secondly, one
could combine the $F$ (inclusive) method with the CPES method.

The magnitude of the time dependent CP asymmetry for various 
final states can be seen in the expression for $Z$ in
Eqn.~(\ref{XYZ_incl}). If 
%$F$ has strangeness=$-1$ so that 
$D^0\to F$ is
Cabibbo allowed 
%and ${\xba D}^0\to F$ is doubly-Cabibbo-suppressed
then $\fd\approx\sin^2\theta_c\approx 0.05$ while $\fb\approx
|V_{ub}||V_{cs}|/(|V_{cb}||V_{us}|)\approx 0.36$. The dominant term in
$Z$ is thus the fourth term which will lead to CP violation of $\lsim
36\%$. Note that this term does not become small in the limit $R_F\to
0$. In the case where $F$ is a CP eigenstate, so $\fd=1$, the second term
$\propto\sin 2\beta\approx 0.8$ becomes dominant. If $R_F$ is small then
the third and fourth terms are dominant, giving again a contribution 
$\propto\fb\approx 0.36$.

Now, in order to illustrate the relative power of the different methods,
let us consider the following toy model.  First, we
%let us 
estimate 
%the color
%suppressed branching ratio $B^0\to\xba D^0 K_S$ to be 
$BR(B^0\to\xba D^0
K_S) \approx \sin^2\theta_c (1/N_c^2) BR(B^0\to D^-\pi^+)/2 \approx
10^{-5}$. For this example, we will arbitrarily take $\eb=50^\circ$ and
$\ed=70^\circ$ with $\gamma=60^\circ$ and $\beta=25^\circ$, consistent
with the B factory values and so $\delta=110^\circ$.

\begin{figure}
\epsfxsize 2.9 in
\mbox{\epsfbox{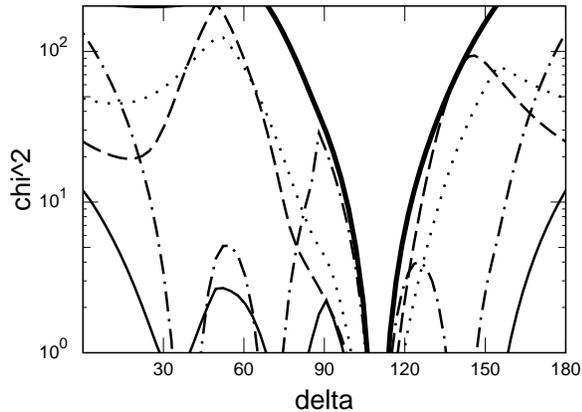}}
\caption{
The $\chi^2_{min}$ vs. $\delta$ for the toy model calculation given $\hat
N_B=10^9$.  The thin solid line is the result for $D^0\to K^-\pi^+$
alone.  The dashed line the result for CPES containing $K_S$ together
with related CPES containing $K_L$.  The dotted lines the result obtained
combining $K^-\pi^+$ with CPES containing $K_S$. The dashed-dotted line
gives the result for $K^-+X$ alone and the thick solid line combines
$K^-+X$ with CPES containing $K_S$. (Note 
the true value of $\delta=110^\circ$)
}
\end{figure}

In Fig.~1 we plot the $\chi^2$ which would be obtained assuming that
${\hat N}_B$
=({\it number of B mesons})({\it acceptance})=
$10^9$ for various
combinations of the data.  The thin solid line gives the minimum value of
$\chi^2$ as a function of $\delta$ obtained for the data from the single
final state $K^-\pi^+$. 
Clearly discrete ambiguities in the solution tend
to conspire to keep the value of $\chi^2$ relatively low.  The dashed line
gives the minimum $\chi^2$ using CP eigenstates containing a $K_S$ and a
$K_L$.  The dotted line shows the case when
$K^-\pi^+$ data is combined with CPES containing $K_S$.
The dashed-dotted line gives
the result for the inclusive $K^-+X$ alone where we have taken for the
purposes of illustration $\hated=70^\circ$, $\hatfd$ the same as for the
$K^-\pi^+$ and $R_F=0.1$ and the thick solid line combines this with the
CP=$-1$ eigenstates. In all cases, we have assumed that the overall
tagging efficiency for the $B^0$ flavor is 25\%.

Table~1 shows the one sigma error on $\delta$ for various inputs.  
Clearly, the best results are obtained when the observables 
overdetermine the parameters and a large fraction of the BR is included in
the sample. Thus when $K^-+X$ is used together with $K_S$ CPES, error on
$\delta$ is $\pm 2.5^\circ$ with ${\hat N}_B=10^9$; with ${\hat
N}_B=10^8$ this error increases to $\pm 11.4^\circ$, so is still
quite useful.

\begin{table}
\begin{tabular}{|p{2.6 in}|l|}
\hline
Case  								&
Accuracy \\
\hline
CPES with $K_S$ and with $K_L$ 					&
$\pm 8.5^\circ$ \\
\hline
CPNES $K^- \pi^+$ with $K_S$ and with $K_L$ 			&
$\pm 5^\circ$\\
\hline
The
CPNES $K^- \pi^+$ together with CPES, both with $K_S$ only   	& 
$\pm 9.0^\circ$ \\
\hline
$K^- + X$ together with $K_S$ CPES 				&	
$\pm 2.5^\circ$ \\
\hline
$K^- + X$ together with $K_S$ as well as $K_L$ CPES		&
$\pm 2.4^\circ$\\
\hline
\end{tabular}
\caption{\emph Attainable one sigma accuracy with various data sets given
${\hat N}_B=10^9$; 
note the 2nd and 5th cases are omitted from Fig~1
for clarity. 
}
\end{table}

Since overdetermining the system of equations is key to improving accuracy
on $\delta$ it may be useful to introduce additional constraints.  First
of all, one can replace the $D^0$ and the $K_S$ with higher resonances
which will tend to increase the total statistics.  In addition, as
suggested in~\cite{kayser} if the $D^0$ is replaced with a $D^{0**}$ then
we can tag the flavor of the $D^{**}$ through the decay $D^{0**}\to
D^+\pi^-$. The analysis of decays with this tag thus reduces to that
of~\cite{branco,sanda}. There is the practical 
problem implementing this method
that one must separate the $D_1$ and $D_2$ states which are 40~MeV apart.
Secondly, as suggested in~\cite{atwood}, using the methods
of~\cite{ggr,in_progress} one can directly determine $R_F$ and $\hated$
from studies at a $\psi(3770)$ charm factory.

Finally, the technique discussed here for replacing a single state with an
inclusive one in the interference of two amplitudes has an immediate
application to getting $\gamma$.
In~\cite{in_progress} we consider this for the method of
~\cite{atwood} for extracting $\gamma$ from $B^-\to K^-D^0$ with various
$D^0$ final states. In particular, this gives a model independent way of
analyzing three body $B$ and $D$ final states.

In conclusion, we show that the ability to determine $\delta$ through
$B^0\to D^0 K_S$ can be greatly enhanced by considering $D^0$ decays to CP
eigenstates and by using inclusive sets of $D^0$ decays. In particular,
using inclusive $D^0$ decays such as $K^-+X$ together with CP
eigenstates, our illustrative calculation suggests that as the number of
available $B$ mesons increases from $10^8$ to $10^9$ a determination of
$\delta$ with a one sigma error of $\lsim \pm 11.4^\circ$ becomes feasible
even with a modest acceptance of $O(10\%)$. This error
can, of course, be reduced to the level of a few percent
as the acceptance is improved. The method described to make
use of inclusive states is likely to have wider application to the
extraction of CP violating phases.

\medskip

This research was supported by Contract Nos.\
DE-FG02-94ER40817 and DE-AC02-98CH10886.


\medskip













\begin{thebibliography}{99}



\bibitem{higuchi} T. Higuchi (BELLE Collaboration) hep-ex/0205020.  %1

\bibitem{raven} G. Raven (BABAR Collaboration) hep-ex/0205045.   %2

\bibitem{ciuch} M. Ciuchini {\it et al}. hep-ph/0012308; D. Atwood and
A. Soni, hep-ph/0103197; A. Hocker {\it et al}., hep-ph/0104062.  %3

\bibitem{cabib} N. Cabibbo, Phys.\ Rev.\ Lett.\ {\bf10}, 531 (1963); M.
Kobayashi and T. Maskawa, Prog.\ Th.\ Phys.\ {\bf49}, 652 (1973).  %4

\bibitem{gron} M. Gronau and D. London, Phys.\ Rev.\ Lett.\ {\bf65},
3381 (1990).  %5

\bibitem{snyder} E. Snyder and H.R. Quinn, Phys.\ Rev.\ D {\bf48}, 2139
(1993); H.R. Quinn and J.P. Silva, Phys.\ Rev.\ D {\bf62}, 054002
(1990).  %6



\bibitem {bigi_sanda} I. Y. Bigi and A. I. Sanda, 
Nucl.Phys. {\bf B193}, 85,1981. 


\bibitem {grontwo} M. Gronau and D. London, Phys.\ Lett.\ {\bf253 B},
483 (1991).   %7

\bibitem {branco} G. Branco, L. Lavoura, J. Silva in
{\it CP Violation}, Oxford Univ. Press (1999), esp. p.445. 


\bibitem{kayser} B. Kayser and D. London, hep-ph/9905561.  %8

\bibitem{sanda} A.I. Sanda, hep-ph/0108031.  %9

\bibitem{dunietz} For a related method using $B^0 \to D^{*+}\pi^-$-like
modes, see,
I. Dunietz, Phys.\ Lett.\ {\bf427B},
179 (1998). Though this method is also clean, as Dunietz 
emphasizes, it leads to very small [O(a few \%)] asymmetry
and seems to require larger B-samples\cite{kayser} than the 
$D^0 K^0$ based methods. %10 

\bibitem{atwood} D. Atwood, I. Dunietz and A. Soni, Phys.\ Rev.\ Lett.\
{\bf78}, 3257 (1997); Phys.\ Rev.\ D {\bf63}, 036005 (2001).  %11


\bibitem{index_note}
For simplicity we use a discrete index for both cases
although a continuous one would be more appropriate if $f_i$ ranges
over
the points in some phase space.


\bibitem{wolf}
L. Wolfenstein, Phys. Rev. Lett. 51, 1945, 1983.


\bibitem{CPK_note}
We also disregard the CP-violation of size $\epsilon_K$ in the $K^0$
system.


\bibitem{pdb} D.~E.~Groom {\it et al.},  
Eur.\ Phys.\ J.\ C {\bf 15}, 1 (2000).  %11


\bibitem{ggr}
M.~Gronau {\it et al.}, hep-ph/0103110.  
%Phys.\ Lett.\ B {\bf 508}, 37 (2001); 
%A. Soffer,
%hep-ex/9801018; 
J.~P.~Silva and A.~Soffer, hep-ph/9912242. 
%Phys.\ Rev.\ D {\bf 61}, 112001 (2000).



\bibitem{in_progress}
D.~Atwood and A.~Soni, in progress. 


%\bibitem{KL_can_do} BELLE\cite{higuchi} and BABAR\cite{raven}
%have demonstrated that $K_L$ detection with good efficiency is
%feasible.


\end{thebibliography}
\end{document}